\documentclass[12pt]{article}

\usepackage{lineno}
\usepackage{graphicx}
\usepackage{multirow}
\usepackage[labelfont=bf]{caption}
\usepackage{subfigure}
\usepackage{booktabs}
\usepackage{amsfonts}
\usepackage{setspace}
\newcommand{\bA}{\mbox{\bf A}}
\newcommand{\bI}{\mbox{\bf I}}
\newcommand{\bU}{\mbox{\bf U}}
\newcommand{\bV}{\mbox{\bf V}}
\newcommand{\bX}{\mbox{\bf X}}
\newcommand{\bZ}{\mbox{\bf Z}}
\newcommand{\bW}{\mbox{\bf W}}
\newcommand{\bu}{\mbox{\bf u}}
\newcommand{\bv}{\mbox{\bf v}}

\newcommand{\bbeta}{\mbox{\boldmath $\beta$}}
\newcommand{\mr}{\multirow{2}{*}}
\newcommand{\mc}{\multicolumn{2}{c}}

\newcommand{\eg}{{\it e.g.} }
\newcommand{\ie}{{\it i.e.} }
\newcommand{\re}{\mathbb{R}}

\newcommand{\rep}{\mathbb{R}_+}
\newcommand{\Zp}{\mathbb{Z}_+}

\begin{document}
\title{Multivariate Survival Mixed Models for Genetic Analysis of Longevity Traits}

\author{Rafael Pimentel Maia, Per Madsen and 
				Rodrigo Labouriau
				\thanks{Corresponding author. Email:rodrigo.labouriau@agrsci.dk 
			\vspace{12pt}}
			\vspace{1pt}	
			\\ 
		         \vspace{12pt}
				{Department of Molecular Biology and Genetics, Aarhus University}}
\date{March 2013}
\maketitle

\begin{abstract}
A class of multivariate mixed survival models for continuous and discrete time with a complex covariance structure is introduced in a context of quantitative genetic applications. The methods introduced can be used in many applications in quantitative genetics although the discussion presented concentrates on longevity studies. The framework presented allows to combine models based on continuous time with models based on discrete time in a joint analysis. The continuous time models are approximations of the frailty model in which the hazard function will be assumed to be piece-wise constant. The discrete time models used are multivariate variants of the discrete relative risk models. These models allow for regular parametric likelihood-based inference by exploring a coincidence of their likelihood functions and the likelihood functions of suitably defined multivariate generalized linear mixed models. The models include a dispersion parameter, which is essential for obtaining a decomposition of the variance of the trait of interest as a sum of parcels representing the additive genetic effects, environmental effects and unspecified sources of variability; as required in quantitative genetic applications. The methods presented are implemented in such a way that large and complex quantitative genetic data can be analyzed. 
\end{abstract}

 \tableofcontents

\newpage

%%%%%%%%%%%%%%%%%%%%%%%%%%%%%%%%%%%%%
\section{Introduction}
%%%%%%%%%%%%%%%%%%%%%%%%%%%%%%%%%%%%%
Longevity is an important trait often considered in animal breeding programs 
\cite{sout03a, sere06, tarres06, yazdi02, ricardi11, vleck80, hoque80, ducroq88, carav04, sout03b, abdelgader12}. 
Even small changes in the longevity of a population under production might have remarkable economic, welfare and ethics 
consequences \cite{sere06, engblom07}.
Since the study of longevity involves several types of incomplete observation (\eg censoring, truncation, late entry and competing risks), survival and event-history-analysis techniques are typically used 
\cite{cox72, kalb02, andersen93}. 
However, the use of those techniques in the context of quantitative genetics of longevity involves several non-trivial challenges. We will present a model framework here that allows to overcome these challenges. These models will have a structure of means and covariances similar to the gaussian linear mixed models classically used in quantitative genetics. They will allow for a proper representation of quantitative genetic phenomena and for efficient implementations required in practice. We will show that these models extend the class of models currently in use for studying the genetic of longevity.

The first challenge in the use of survival models for characterizing the genetic aspects of longevity is the high complexity of the models. Typically it is necessary to adjust simultaneously for the effects of several explanatory variables, some continuous with linear effects 
(\eg breed composition and heterosis \cite{lopes10}), 
some factors with many levels (\eg herd) or even time-dependent effects (\eg year and season). Further complexity is added by the necessity of representing complex genetic effects (several generations deep pedigrees) under different genetic models (\eg sire model, sire-dam-model etc). A typical scenario of applications in quantitative genetics involves a very large number of observations (individuals); indeed often the analyses involve several hundred thousand of observations (we will present two relative small illustrative examples involving 142,133 and 200,084 individuals). This leads to inference problems of the complexity equivalent to solve systems of linear equations with several hundred of thousand, or even several million, simultaneous linear equations (in our examples between 8,248 and 118,432 simultaneous linear equations) when using classic gaussian linear mixed models. Now, the complexity of survival and event-history models is even larger since these models require to keep track of the individuals' history. Moreover, as it will be apparent from the description of the methods available and from our discussion, some of those models might present very flat likelihood functions making the inference problem hard and numerically unstable. This rules out the naive use of the standard methods of survival analysis and makes the use of specially constructed models and approximations mandatory. 

A second challenge is the proper representation of the genetic phenomena in play. The interpretation of the gaussian linear mixed models in genetic terms requires a linear decomposition of the so called total variance (\ie the phenotypic variance after having corrected for the effects of a range of non-genetic related aspects with known effects on the trait of interest)\cite{jacquard83, viss08}.
This decomposition of the total variance occurs naturally in the gaussian linear mixed models but not in all the adaptations of survival models specially constructed for genetic evaluations of longevity.
We will show that a way to circumvent this problem is to consider models containing a dispersion parameter which plays the role of the residual variance (\ie the part of the total variance that is neither attributed to genetic components nor to identifiable sources of variation) and base the inference on the quasi-likelihood theory \cite{wedd74, bresl93}. The use of generalised linear mixed models with a dispersion parameter occurs only occasionally in the literature of quantitative genetics 
\cite{guerra04, butler04}, and the genetics and the statistics consequences of using a free dispersion parameter has not being systematically explored yet. Moreover, survival models with dispersion parameter were, to the best of our knowledge, never considered in the literature of quantitative genetics.

A third challenge is the presence of incomplete observations following complex patterns. Variants of the so called Cox proportional model \cite{cox72} were used for univariate traits describing longevity 
\cite{ducroq88, giolo11, ducroq94}.
These techniques succeeded to implement models of approximations that were operational for some of the purposes of animal breeding. However, all those models were of univariate nature (\ie they consider one trait at time)
\cite{sere06, ducroq99},  
which makes it difficult, if not impossible, to address some key questions related to the issues of competing risks, informative censoring and inhomogeneity of the failure mechanism. A clear example is the study of time to death were the individuals on the study may die from one among several different causes of death. Moreover, the use of multivariate models is necessary to study the time death in a competing risk scenario, by evaluating the risk of death for each specific cause simultaneously. This allows to form the basis to study further aspects as the presence of common genetic determining factors and/or independent or specific genetic factors for each of the death causes. In the framework presented here, it will be possible to treat these issues by using suitable multivariate models.

Two types of models will be considered: models based on continuous time and models based on discrete time. The continuous time models (CTM) will be suitable approximations of the semi-parametric Cox proportional model in which the hazard function will be assumed to be piece-wise constant. These models allow for regular parametric likelihood-based inference by exploring a coincidence of their likelihood functions and the likelihood function of a Poisson model applied to a specially constructed data \cite{aitk80}. When including gaussian random components (log gaussian frailties) the inference can be based on a Laplace approximation to a high dimensional integral \cite{bresl93, madsen10b, ripatti00}. Here, we will include also a dispersion parameter that will play the role of the residual variance which is an essential element for the genetic interpretation of the models. The CTM will depend on an arbitrary choice of time cut points used to define the intervals at which the hazard function is assumed to be constant.

The discrete time models (DTM) used are multivariate variants of the classic discrete relative risk models \cite{kalb02}. These models present advantages in terms of computational and statistical complexity as compared to the CTM: the algorithms run faster relatively to the analogous algorithm for CTMs, are numerically more stable, and the statistical inference is more efficient. The loss of genetic information occurred when switching from the CTM to the DTM is minimal in typical applications in quantitative genetics and this loss is fairly compensated by the avoidance of capturing a large amount of noise.  Here, we will also introduce the use of a dispersion parameter, which will be essential for well representing genetic scenarios.

The aim of this article is to introduce, characterize and discuss a class of multivariate mixed survival models for continuous and discrete time with a complex covariance structure in a context of quantitative genetics applications. The methods introduced here can be used in many other applications in quantitative genetics although the discussion presented concentrates on longevity.

The paper is organized as follows. Section \ref{Sect.1} describes the basic set-up and the genetic scenario discussed. There, a suitable multivariate version of the proportional hazard model is introduced in general terms. Those models will encompass models for competing risks possibly defined with different types of time scale (continuous and discrete time). The techniques for statistical inference under those models are presented in section \ref{Sect.2}, including  a general discussion on the calculations involved in the likelihood based inference and some connections with multivariate generalized linear mixed models (section \ref{Sect.2.1}). A Poisson approximation useful for efficiently implementing likelihood based inference for continuous time models is presented in section \ref{Sect.2.2}.  Section \ref{Sect.2.3} presents an extension of the models to a situation where there is a stratification variable. A discussion comparing the amount of statistical  information between models constructed with discrete time and models based on continuous time is given in section \ref{Sect.2.4}. The quantitative genetic theory behind the models considered here requires a special decomposition of the phenotypic variance in terms of the variance of the random components and a scale parameter present in the models. This will be necessary for the calculation of the so called heritability, which is  crucial in quantitative genetic applications. Section \ref{Sect.3} will discuss methods for that and some technical details involving counting processes and martingale theory are presented in the appendix. Two illustrative examples involving longevity of sows and dairy cattle will be presented in section \ref{Sect.4.1} and \ref{Sect.4.2} respectively. Some discussion will be given in section \ref{Sect.5}. 

%%%%%%%%%%%%%%%%%%%%%%%%%%%%%%%%%%%%%
\section{The basic set-up and genetic scenario}
\label{Sect.1}
%%%%%%%%%%%%%%%%%%%%%%%%%%%%%%%%%%%%%

The class of models we will discuss are thought to be applied in the following  general scenario. The life spans of a population of $n$ individuals are observed. Genetic information (typically in the form of a reasonably deep pedigree or comprehensive genomic data) and  the values of $k$ explanatory variables are available for each individual. The interest lies in modeling the longevity, \ie the length of the life span or the length of the productive life of the individuals, with particular interest on some forms of genetic determination typically used in quantitative genetics.

The longevity can be operationally measured using a continuous time, by determining the time elapsed between two life events in the life of the individuals (\eg  the time elapsed from the first parity the to the culling of cows under production) or a discrete time, by counting a number of certain events during the life of the individual (\eg by counting the number of survived parities of cows). We will present two illustrative examples: one involving the longevity of sows and the other studying the longevity of dairy cattle. Many similar examples can be found in the literature  with other husbandry animals as ewes \cite{maia2013}, salmon \cite{labouriau2011} among others.

An additional issue that might occur in longevity studies is the presence of competing risks, which arises when the individuals could be culled for one of two or more types of culling. For instance, in the example of dairy cattle longevity, the cows might die or be slaughtered, and the interest is in studying the culling rates for both causes. We assume in this scenario that there are two causes of death or culling, which will be labeled by the index $j$. It is straightforward to extend the setup described here to the general case with more than two culling causes or to the case with only one culling type.

The longevity will be characterized by the time development of the rate at which the individuals die or are culled. This is measured differently when using continuous or discrete time as we define below. In order to describe the models we have in mind, consider two random variables: $T$ representing the time (continuous or discrete) of culling and $J$ indicating the cause (or type) of culling. In the case of the continuous time, we define the {\it cause-specific hazard function} \cite{kalb02} for the $j^{\rm th}$ culling cause (for $j=1,2$) by
\begin{eqnarray*}
 \lambda_{[j]}(t) = 
 \lim_{\Delta \downarrow 0} 
 \frac{ P(t \leq T < t + \Delta , J=j \vert T > t) }
        {\Delta} 
  \;\;\; , \; \forall \, t\ge 0 
 \, .
\end{eqnarray*}
In contrast, when the time is discrete, we define the {\it cause-specific hazard probability function} \cite{kalb02} for the $j^{\rm th}$ culling cause (for $j=1,2$) by
\begin{eqnarray*}
 \lambda_{[j]}(t) = P(T = t , J=j \vert T \geq t)  \mbox{ for } t=0,1,2, \dots \; .
\end{eqnarray*}
Although these two characterisations of the time development of the culling rate are of different nature, they will essentially play the same rule in the models discussed 
and will be generically referred simply as the cause-specific hazard function.
Note that in the case where only one cause of culling is studied the cause-specific hazard function and the cause-specific hazard probability function defined above coincide with the hazard function and the hazard probability function used in the literature of survival analysis (see \cite{kalb02}).

The models presented below will assume a scenario where there are two causes of culling (indexed by $j$). Here, the cause-specific hazard functions (or the hazard probability functions) are specified in terms of $k$ explanatory variables (the fixed effects) and a set of gaussian random components. Without loss of generality, we consider only two random components for each culling cause: $\bU_1$ and $\bU_2$ that will represent additive genetic effects (usually determined using information on the pedigree of the animals in play) for cause $1$ and $2$ respectively and  $\bV_1$ and $\bV_2$, representing an environment effect (\eg the effect of herd where an animal is kept) for cause $1$ and $2$, respectively. The random components related to genetics (\ie $\bU_1$ and $\bU_2$) are independent of the components related to environment ($\bV_1$ and $\bV_2$) by construction. This constraint can be relaxed for describing possible interactions between environmental and genetic effects, but we will not pursue this project here.
The extension of the present definition to a situation with a different number of random components is straightforward. According to the general model we describe, the  vector of cause specific hazard functions of the $i^{\mbox{th}}$ individual (for $i=1, \dots, n$), conditionally on $\bU_1 = \bu_1$, $\bU_2 = \bu_2$, $\bV_1 = \bv_1$ and $\bV_2 = \bv_2$ (for the realizations $\bu_1$ , $\bu_2$, $\bv_1$ and $\bv_2$  of $\bU_1$ , $\bU_2$, $\bV_2$ and $\bV_2$, respectively),  is given by 
\begin{equation} \label{eqn.1.1}
  \left[ \begin{array}{c}
	    \lambda_{[1],i}(t_1 \vert  \bU_1,  \bV_1) \\
	    \lambda_{[2],i}(t_2 \vert  \bU_2, \bV_2)	    
	  \end{array}
  \right ]	
  = 
  \left[ \begin{array}{c}	
	\lambda_{1}(t_1) \exp 
	\left ( \bX_{1i}^{'}(t_1)\bbeta_1 + \bZ_{1i}^{'}\bu_1 + \bW_{1i}^{'}\bv_1 \right )
	\\
	\lambda_{2}(t_2) \exp 
	\left ( \bX_{2i}^{'}(t_2)\bbeta_2 + \bZ_{2i}^{'}\bu_2 + \bW_{2i}^{'}\bv_2 \right )

  \end{array}
  \right ]	
	 \; .
	\label{pchm}
\end{equation}	
Here the times $t_1$ and $t_2$ might be continuous or discrete; moreover, $t_1$ and $t_2$ are not necessarily of the same type (as in the example of sows longevity considered in section \ref{Sect.4.1}). Equation (\ref{eqn.1.1}) holds for  $t_j\ge 0$ when $t_j$ is continuous and for $t_j= 0,1,2,\dots $ if $t_j$ is discrete ($j=1,2$). Furthermore,  $\lambda_1(\cdot)$ and $\lambda_2(\cdot)$ are baseline hazard functions describing a common time development of the cause specific hazard functions (typically considered as a nuisance parameter), $\bbeta_1,  \bbeta_2 \in \re^{k}$ are (finite dimensional) parameters representing the fixed effects (also viewed as a nuisance parameter),  $\bX_{i}$, $\bZ_{ji}$ and 
$\bW_{ji}$ are incidence matrices for the fixed, genetic and environment effects for the $j^{\mbox{th}}$ cause and the $i^{\rm th}$ individual. When the time is continuous $\lambda_{[j],i} (\cdot)$ in (\ref{eqn.1.1}) is a hazard function and the marginal model described will correspond to a variant of the Cox proportional  hazard model with frailties \cite{kalb02,andersen93}. If the time is discrete, $\lambda_{[1],i}$ in (\ref{eqn.1.1}) is a hazard probability function and the marginal model described will be a variant of the discrete relative risk model with frailties \cite{kalb02}. 

In the case of the models based on continuous time, we assume that the baseline function, $\lambda_j(\cdot)$ is piece-wise constant. That is, we assume that there is a set of disjoint intervals, $I_1, \dots , I_K$ covering the positive real line, in which $\lambda_j(\cdot)$ is constant. This constraint in the class of possible baseline functions will allow us to connect the models described here with some generalized linear mixed models: Moreover, this will allow us to perform efficient inference with complex data using existent software.

The structure of covariance of the random components is a crucial part of the construction of the models for longevity described since this is the part of the model where the additive genetics and the environment contributions to the cause specific hazard functions are considered.
We assume the random components to be multivariate normally distributed, \ie $(\bU_1, \; \bU_2, \; \bV_1, \; \bV_2)^{'} \sim \mathcal{N} \left( {\bf 0}, {\bf \Sigma} \right )$ where
\begin{equation} \label{eqn.1.2}
{\bf \Sigma} =
\left[ \begin{array}{cc} 
         \Sigma_1 \otimes \bA & \bf 0 \\ \bf 0 & \Sigma_2 \otimes \bI
      \end{array} 
\right] 
\; .
\end{equation}	
Here, ${\bf I}$ is an identity matrix and ${\bf A}$ is the $n\times n$ additive relationship matrix constructed as follows.  Each entry of ${\bf A}$ represents a pair of individuals in the population. $A_{ii} = 1$ for $i=1, \dots , n$, and  $A_{ij} = 0$ if the pair of individuals $(i,j)$ are not related. In the case were the pair of individuals $(i,j)$ are related, we consider the  
degree of relationship  $r$ given by 1 for relatives of first-degree, 2 for relatives of second-degree, 3 for relatives of third-degree and so on. In that case $A_{ij} = (1/2)^r$. The additive relationship between two individuals measures the proportion of identical by descent genes expected to be shared by pairs of individuals in the population. The methods introduced here could also be applied to genomic analyses where the relationship matrices are inferred from genomic data.

The covariance structure between the genetic random components is given by
$$
\Sigma_1 = \left[ \begin{array}{cc} \sigma_{g.1}^2 & \sigma_{g.12} \\ \sigma_{g.12} & \sigma_{g.2}^2 \end{array} \right]
\; .
$$
In particular the genetic correlation between the two culling causes is given by 
$\rho_g = \frac{\sigma_{g.12}}{\sqrt{\sigma_{g.1}^2 \sigma_{g.2}^2}}$.
Finally, the covariance structure between the environment random components is given by
$$
\Sigma_2 = \left[ \begin{array}{cc} \sigma_{e.1}^2 & \sigma_{e.12} \\ \sigma_{e.12} & \sigma_{e.2}^2 \end{array} \right] 
\; .
$$

%%%%%%%%%%%%%%%%%%%%%%%%%%%%%%%%%%%%%
\section{Inference}
\label{Sect.2}
%%%%%%%%%%%%%%%%%%%%%%%%%%%%%%%%%%%%%

%------------------------------------------------------------------------------------------------
\subsection{Representation of the counting process and connections with generalized linear mixed models}
\label{Sect.2.1}
%------------------------------------------------------------------------------------------------

The statistical inference for the longevity models can be made by using the coincidence of the conditional likelihood of those models (conditioned on the frailties) with the conditional likelihood of a generalized linear model applied to a specially designed pseudo data representing an underlying counting process as briefly described below. In the case of the continuous time, the pseudo data is constructed by generating one pseudo observation for each time interval (in which the baseline function is assumed to be constant) observed for each individual. In the case of the discrete time, the pseudo data contains one observation per each unit of time observed for each individual. To facilitate the description we use the term "observed times" to refer to the observed discrete times or the time intervals related to the observed continuous time. A pseudo response variable was created for each of the culling causes. These variables are such that they take the value 0 when an individual is at risk of being culled for the current reason but was not culled or take the value 1 if the individual was culled for the current reason in that period. This pseudo data describes the groups of individuals at risk at each observed time.

The statistical inference is performed by using a multivariate generalized linear mixed model (MGLMM) applied to the pseudo data. One dimensional versions of the calculations we describe below were given previously in \cite{aitk80} for models without random components and in \cite{ha03} for mixed models for continuous time models (piece-wise constant baselines). For one-dimensional discrete times models without random components see \cite{kalb02}. Each of the marginal model of this MGLMM corresponds to a culling reason and is defined with a logarithmic link function. The distribution of each marginal model is Bernoulli (or binomial) when the corresponding time is discrete or Poisson when the time is continuous. Furthermore, the marginal models constructed with continuous time include an offset variable with the logarithm of the length of the corresponding time intervals. Additionally, a discrete explanatory variable counting the order of the time period for each individual should be included in the model in order to represent the baseline function.
The model should also specify the covariance structure of the random components given in equation (\ref{eqn.1.2}). Furthermore, each marginal model includes a dispersion parameter $\phi_j$ (known in the literature of generalized linear models as an over-dispersion parameter for binomial and Poisson models), which allowed us to better characterizing the genetic scenario. 

 The MGLMM described above can be adjusted with any software implementing multivariate generalized linear mixed models, which allows the inclusion of over-dispersion parameters in binomial and Poisson models via quasi-likelihood inference \cite{wedd74, bresl93}. We used the software DMU version 6.0, release 5.1 \cite{madsen10a, madsen10b}, which implements multivariate generalized linear mixed models efficiently and facilities to specify the required covariance structure of the random components.
  
%------------------------------------------------------------------------------------------------
\subsection{A Poisson approximation for discrete-time models}
\label{Sect.2.2}
%------------------------------------------------------------------------------------------------

The inference for the discrete time models presented here is equivalent to making inference for a Bernoulli model where the response variable is a variable indicating whether a culling event occurred. Here, we present a Poisson approximation that will be useful for performing likelihood-based inference for the discrete time models, specially in the case of very large populations with relatively low occurrence of culling. 
More precisely, for each time $t$ ($t = 0,1.\dots,\tau$, where $\tau$ is the maximum observed time) and for each individual $i$ ($i = 0,1,\dots,n_t$ where $n_t$ is the number of individuals at risk in the time $t$), we constructed a pseudo observation of a response variable $Y_{it}$ taking the value 1 if the individual $i$ died at time $t$ and 0 otherwise. Exploring a coincidence of the likelihood function, we treated the observations $Y_{it}$, for $t = 0,1.\dots,\tau$ and $i = 0,1,\dots,n_t$, as independent Bernoulli random variables with $P(Y_{it} = 1) = p_{it}$, where $p_{it}$ is the hazard probability for the individual $i$ at the time $t$. 

The following condition will ensure that the likelihood function of a suitably defined Poisson model will approximate the likelihood function of the Bernoulli models considered here. Suppose that 
\begin{eqnarray} \label{Eq.Poisson.1}
\max_{1 \leq t \leq \tau} \bar{p}_{t,n_t} \rightarrow 0 \,\, ,
\end{eqnarray}
where $\bar{p}_{t,n_t} = \max_{1 \leq i \leq n_t} p_{it}$, and
\begin{eqnarray} \label{Eq.Poisson.2}
 \sum_{i=1}^{n_t} p_{it} = \gamma_t \mbox{ (fixed) } \mbox{ for all } t 
			\mbox{ as } \min_{1 \leq t \leq \tau} n_t \rightarrow \infty
\,\, .
\end{eqnarray}

The general result on convergence for arrays of probabilities presented in \cite{chen1974} implies that, under the conditions (\ref{Eq.Poisson.1})  and (\ref{Eq.Poisson.2}), for each time $t$ and $r=0,1, \dots$,
\begin{equation}
\lim_{n_t \rightarrow 0} P(W_{n_t} = r) = \frac{e^{\gamma_t} \gamma_t^r}{r!}
\,\, ,
\label{Eq.3.2}
\end{equation}
where $W_{n_t} = \sum_{i=1}^{n_t} Y_{it}$.  Note that the right side of the equation (\ref{Eq.3.2}) is the density probability function of a Poisson random variable with expected value equal to $\gamma_t$. In other words, if the number of individuals at risk in each time $t$ (\ie $n_t$) is larger enough and the probability of death in each time $t$ is small enough, then the likelihood function of the Bernoulli model is approximated by the likelihood function of a Poisson model.

Note that the conditions given in (\ref{Eq.Poisson.1})  and (\ref{Eq.Poisson.2}) are reasonable in the example presented here, since the number of individuals is of the order of 100 thousands at the first observed period and is of the order of 16 thousands at the last observed period. Moreover, for a given individual $i$, the pseudo random variables $Y_{it}$ ($t = 1,\dots,t_i$) is a sequence of zeros until time $t_i - 1$ and takes the value1 at $t_i$ only if this individual was culled, therefore $p_{it}$ will be small in general.

%------------------------------------------------------------------------------------------------
\subsection{Continuous models with stratification}
\label{Sect.2.3}
%------------------------------------------------------------------------------------------------

The examples of survival models with continuous time presented here required the use of stratification of the baseline function, which is a standard technique of survival analysis. For example, we used the number of days from the first parity to the culling day as one of the characterizations of the longevity of animals (sows or dairy cows), in the two examples presented here. Exploratory analyses of the data used indicated that there were differences in the mortality rates between the different parities and that the form of the baseline functions were not the same for the different parities, \ie the baseline functions for the different parities were not proportional. A way to circumvent this issue is to use the stratification technique defined here. This type of model is referred in the literature of dairy cattle longevity as the "lactation basis model" (see  \cite{roxstr03}).

Define $t_{p_i}$ as the time of the occurrence of the $p^{\rm th}$ parity of the individual $i$ ($i = 1,\dots, n$ and $p = 1,\dots,P_i$ where  $P_i$ is the maximum parity observed for the individual $i$) and $t_i^*$ as the observed time (death or censure) for the $i^{\rm th}$ individual. The hazard function for the $i^{\rm th}$ individual at the $p^{\rm th}$ parity, for $i = 1,2,\dots, n$ and for $p = 1,\dots,P_{i}$ conditional on the random components $\bU = \bu$ and $\bV = \bv$ is then
\begin{eqnarray}\nonumber
\lambda_{i,p}(t \vert \bU, \bV) = Y_{i,p}(t) \lambda_p (t) \exp \left( \bX_{i,p}\bbeta + \bZ_i \bu + {\bf W}_i \bv \right),
\label{eq.5.1}
\end{eqnarray}
for $t \in [t_{p_{i}}, t_{\left({p_i+1} \right)})$ when $p < P_i$ and for $t \in [t_{P_i},t_i^*)$ when $p = P_i$. Here $\lambda_p (\cdot)$ is the baseline hazard function for the  $p^{\rm th}$ parity. The variable  $Y_{i,p}(t)$ is equal to 1 if the individual was at risk at $t \in [t_{p_{i}}, t_{\left({p_i+1} \right)})$ and $0$ otherwise.
In this study, the time dependent explanatory variables will be a function of $p$ and then $X_{i,p}(t) = X_{i,p}$ for all $t > 0$. 

%------------------------------------------------------------------------------------------------
\subsection{Comparison of the statistical information between discrete and continuous time models}
\label{Sect.2.4}
%------------------------------------------------------------------------------------------------

When applying the longevity models described above it is often possible to choose between models defined with continuous time and models defined with discrete time. For instance, in the examples studied in section \ref{Sect.4} the longevity can be characterized by the time between the first parity and the culling (continuous time) using a stratification by parity or, alternatively, by the number of survived parities (discrete time). The complexity and the statistical stability of these two types of models differ, which is an important aspect to take into account when deciding which type of model to use. We will argue next that the  likelihood function of a continuous time model is typically much flatter than the likelihood function of a corresponding discrete time model. This explains certain anomalies in the inference under continuous time models.

In order to simplify the exposition, we assume a scenario where two models are considered: a discrete time model where the number of survived parities is modeled and a continuous time model where the number of survived days is modeled but there is a stratification by parity (as described in section \ref{Sect.2.3}). Moreover, we restrict the discussion to the case where there is only one cause of culling and there is only one random component, say $\bU$, in the model. Here, $\bU$ will be a multivariate normal random variable with $E(\bU) = {\bf 0}$, $Var(\bU) = \bA \sigma^2$, where $\bA$ is a known matrix. We denote the density function of the distribution of $\bU$ by $ \Phi \left( \cdot, \sigma^2 \right)$ and represent the multiple integral for integration with respect to this density function by a single integration sign. 

The likelihood function of the piece-wise constant hazard model stratified by parity based on the number of survival days is 
\begin{eqnarray} \label{Eq.Lik.PCHM}
 \int \prod_{i,p,k} \exp \left\{ - \Delta_{ipk} \exp \left( \eta_{ipk} \right) \right\} \left\{ \Delta_{ipk}\exp \left( \eta_{ipk} \right) \right\}^{y_{ipk}} \Phi \left( \bu, \sigma^2 \right) d\bu \, \, , 
\end{eqnarray}
where the indices $i,p$ and $k$ indexes the individual, the parity and the time intervals in each periods, respectively. Here, $y_{ipk}$ is a indicator variable taking the value 1 if the $i^{\rm th}$ individual died at the $p^{\rm th}$ parity in the $k^{\rm th}$ time interval and 0 otherwise; $\Delta_{ipk}$ is the length of the $k^{\rm th}$ time interval at the $p^{\rm th}$ parity; and $\eta_{ipk} = \log(\lambda_{pk}) + \bX_{ipk}^{'} \bbeta + \bZ_{i}^{'}\bu$ is the linear predictor. 

On the other hand, by the Poisson approximation discussed in section \ref{Sect.2.3}, the likelihood function of the discrete relative risk model based on the number of survived parities is approximated by 
\begin{eqnarray} \label{Eq.Lik.DRRM}
 \int \prod_{i,p} \exp \left\{ - \exp \left( \eta_{ip} \right) \right\} \left\{ \exp \left( \eta_{ip}  \right) \right\}^{y_{ip}} \Phi \left( \bu, \sigma^2 \right) d\bu \, \, , 
\end{eqnarray}
where $i$ and $p$ indexes the individual and the parity, respectively; $y_{ip}$ is the indicator variable that the $i^{\rm th}$ individual died at the $p^{\rm th}$ parity and $\eta_{ip} = \log(\lambda_p) + \bX_{ip}^{'} \bbeta + \bZ_{i}^{'}\bu$ is the linear predictor.

Note that, when there are many censured observations and when there are many survived parities or many observed time intervals in each parity, the variables $y_{ipk}$ and $y_{ip}$ will take the value zero in most of the cases. Therefore, under these conditions, the integrand in (\ref{Eq.Lik.PCHM}) will be dominated by 
\begin{eqnarray}
\label{Eq.Lik.01}
\prod_{i,p,k} \exp \left\{ - \Delta_{ipk} \exp \left( \eta_{ipk} \right) \right\} 
= \exp \left\{ - \sum_{i,p,k} \Delta_{ipk} \exp \left( \eta_{ipk} \right) \right\} 
\end{eqnarray}
and the integrand in (\ref{Eq.Lik.DRRM}) will be dominated by 
\begin{eqnarray}
\label{Eq.Lik.02}
\prod_{i,p} \exp \left\{ - \exp \left( \eta_{ip} \right) \right\} 
= 
\exp \left\{ - \sum_{i,p}\exp \left( \eta_{ip} \right) \right\}
\,\, .
\end{eqnarray}
Note that the sum in the right side of (\ref{Eq.Lik.01}) has typically a larger number of parcels as compared to the sum in the right side of (\ref{Eq.Lik.02}). Moreover, the parcels in the sum in the right side of (\ref{Eq.Lik.01}) are multiplied by the factors $\Delta_{ipk}$ which are typically larger than 1. Note, moreover, that if one reduces the size of the intervals used in the piecewise constant hazard model, the factors $\Delta_{ipk}$ decrease, but the number of parcels in the sum in the right side of (\ref{Eq.Lik.01}) increases; on the other hand, if one decreases the number of intervals, the number of parcels decreases, but the factors $\Delta_{ipk}$ might increase substantially. This makes the likelihood function of the piecewise constant hazard models much flatter, since the likelihood function is positive. More precisely,  the curvature of the support curve (\ie the graph of the logarithm of the likelihood function) near the maximum likelihood estimate will be much smaller for the piecewise constant hazard models. Therefore, the Fisher information will be smaller, and the maximization  of the likelihood function will be a much harder problem, for the piecewise constant hazard models as compared to the corresponding discrete relative risk model.
 
%%%%%%%%%%%%%%%%%%%%%%%%%%%%%%%%%%%%%
\section{Decomposition of the phenotypic variance and heritability}
\label{Sect.3}
%%%%%%%%%%%%%%%%%%%%%%%%%%%%%%%%%%%%%

The notion of heritability is crucial in quantitative genetic applications; it measures the magnitude of the detected additive genetic signal relative to the total variance of a trait of interest and is typically used to quantify the potential response to selection. Heritability  (in the narrow sense) is defined operationally by the ratio 
\begin{equation}
h^2 = \frac{ \sigma_g^2 }{ \sigma_P^2 } 
\; ,
\label{Eq.4.1}
\end{equation}
where $\sigma_g^2$ is  the variance of an unobserved  random component representing all the additive genetic variation (termed the genetic variance) and  $\sigma_p^2$ is the variance of the trait of interest (called the phenotypic variance). For details see \cite{lynch1998, jacquard83}. In the classic scenario, where gaussian linear mixed models are used, the calculation of the heritability is straightforward because the phenotypic variance can be expressed as the sum of the genetic variance $\sigma_g^2$ and the variances of the other random components present in the model. However, the calculation of the heritability is much more complicated in the context of survival analysis considered here and are, therefore, developed in full details below. It will be necessary to introduce some details of the counting process behind the models used and to define the trait of interest carefully. Here, we present the main results informally and expose the precise formal definitions and the technical details using the mathematical machinery of counting processes in appendix \ref{ref.appendix.01}. For simplicity of exposition, we discuss only the one-dimensional case were only one cause of culling is present (referred as death) and assume the same model structure presented in section \ref{Sect.2} but with dimension one.

Let $T$ be a non-negative random variable representing the observed survival time and $D$ be an indicator variable taking the value 1 if a death is observed and 0 otherwise (censure). The random variable $T$ will take values in $\rep$ if the time is continuous or in $\Zp$ (\ie in $\{ 0, 1, 2, \dots \}$) if the time is discrete. We assume that for each time $t$ a vector $X(t)$ of explanatory variables (possibly changing with time) is known. Denote by $\bX(t) = \{ X(s): 0 \leq s < t \}$ the trajectory (or path) function until the time just before $t$ (denoted by $t-$). Additionally, we consider two independent gaussian random components, $\bU$ and $\bV$, representing the additive genetic effects and some environmental effects, respectively. Here, we follow the same convention as in section 2, so  $Var(\bU) = \bA \sigma_g^2$ and $Var(\bV) = \bI \sigma_e^2$, where $\bA$ is the relationship matrix and $\bI$ is an identity matrix with suitable dimensions. The hazard function for the $i^{\rm th}$ individual takes then the form
\begin{equation} \label{Eq.4.2}
	    \lambda_{i}(t \vert  \bU,  \bV)
	     = 	
	    \lambda_{0}(t) \exp \left ( \bX_{i}^{'}(t)\bbeta + \bZ_{i}^{'}\bu + \bW_{i}^{'}\bv \right )
	    \, .
\end{equation}	

In order to describe the models presented in section \ref{Sect.2} properly, we need to introduce two stochastic processes: the death (or event) counting process and the at risk process. To define the death process we consider, for each $t$, the random variable $N(t)$, which indicates whether a death occurred until the time $t$, \ie $N(t) = \mathbf{1}(T \leq t, D = 1)$. The stochastic process $\mathbf{N} = \left \{ N(t) : t \in \rep \mbox{ or } \Zp \right \}$ is the death counting process. It is convenient to introduce the notation $dN(t)$ to denote the increments of the counting process $\mathbf{N}$ at each time $t$, \ie $dN(t) = N(t) - N(t-)$, where $N(t-)= \lim_{ \Delta \downarrow 0} N(t - \Delta)$. Furthermore, the at risk process given by $\mathbf{Y} = \left \{ Y(t) : t \in \rep \mbox{ or } \Zp \right \}$, where, for each time $t$, the random variable $Y(t)$ takes the value $1$ when the individual is at risk at time $t$ or $0$ otherwise.  

We will calculate the phenotypic variance and the heritability for two characteristics of interest: the hazard function evaluated at a given time and the accumulated hazard function up to a given time. These two characteristics are associated to the intensity and the cumulative intensity of the counting process of death $\mathbf{N}$. Due to the structure of the models used here (see equation (\ref{eqn.1.1}) and (\ref{Eq.4.2}) for instance), the entire hazard function of an individual is multiplied by the factor $\exp (\bu)$, where $\bu$ is the realization of the genetic random component $\bU$ for that individual; so, according to this model, when selecting individuals by the predicted values of $\bU$ (as it is the usual practice in animal breeding), the entire hazard curve is affected and therefore, it is indifferent if one measures the potential response to selection (\ie the heritability) of the hazard evaluated at a given time-point, say $t$, or even of the accumulated hazard up to $t$. 

In order to study the heritability related to the hazard function evaluated at a given time $t$ (in $\rep$ or in $\Zp$), we define the conditional random variable 
\begin{equation} \label{eq.4.1}
{\cal{Y}}(t) = dN(t) \, \vert \, T\geq t, X(t) \, .
\end{equation}
Note that $E\left ({\cal{Y}}(t)\, \vert \, \bU = \bu , \, \bV = \bv \right ) = \lambda \left (t \,\vert \, \bU = \bu , \, \bV = \bv \right )$, therefore ${\cal{Y}}(t)$ will be treated as the trait of interest, for which we calculate the phenotypic variance and the heritability. A more precise definition of ${\cal{Y}}(t)$ using counting processes theory is given in appendix \ref{ref.appendix.01}. There we show, taking advantage of the basic theory of martingales and using a suitable Taylor expansion, that in the continuous time case the variance of ${\cal{Y}}(t)$ is approximated by
\begin{equation}
Var \left[ {\cal{Y}}(t) \right] \approx 
Y(t) \left \{
\left[  \lambda^*(t) \right]^2 \left( \sigma_g^2 + \sigma_e^2 \right) + \phi \lambda^*(t)
\right \}
\,\, ,
\label{Eq.4.6}
\end{equation}
where $\lambda^*(t)$ is the hazard function evaluated at $t$ with vanishing random components, \ie
\begin{equation}
\lambda^*(t) = \lambda(t \vert \bU = \bf{0}, \bV = \bf{0}) = \lambda_{0}(t) \exp \left \{ \bX_{i}^{'}(t)\bbeta  \right \}
\, \, .
\label{Eq.4.6B}
\end{equation}
Note that according to (\ref{Eq.4.6}) the variance of  ${\cal{Y}}(t)$ is additively decomposed in three components: one depending on $\sigma^2_g$, one depending on $\sigma^2_e$ and one depending on $\phi$. This is a situation analogous to the classic gaussian linear mixed models. Here, the component of the  variance of  ${\cal{Y}}(t)$ associated to $\sigma^2_g$, namely  $\left[ Y(t)\lambda^*(t)\right]^2 \sigma_g^2$ , plays the role of the genetic variance and therefore the heritability for the hazard evaluated at the time $t$ is given by
\begin{equation}
h_{ \lambda(t) }^2 
\approx  Y(t) \frac{ \sigma_g^2 }{\sigma_g^2 + \sigma_e^2 +  \frac{\phi }{\lambda^*(t)}}
\,\, .
\label{Eq.4.7}
\end{equation}

Analogous calculations (see appendix \ref{ref.appendix.01}) yield for the discrete time case that the variance of  ${\cal{Y}}(t)$ is approximated by 
\begin{equation}
Var \left[ {\cal{Y}}(t) \right] \approx Y(t) \left \{ \left[ \lambda^*(t) \right]^2 \left( \sigma_g^2 + \sigma_e^2 \right) + \phi \lambda^*(t)[1 - \lambda^*(t)]
\right \}
\,\, .
\label{Eq.4.9}
\end{equation}
Here, as in the case of the continuous time, the variance of ${\cal{Y}}(t)$ is additively decomposed in three separate a components depending on  $\sigma^2_g$, on $\sigma^2_e$ and on $\phi$, respectively. Therefore, the heritability is approximated by
\begin{equation}
h_{ \lambda(t) }^2 
\approx  Y(t) \frac{ \sigma_g^2 }
                            { \sigma_g^2 + \sigma_e^2 +  \phi \frac{  \lambda^*(t)}
					                                       {1 - \lambda^*(t)}
          }
\, .
\label{Eq.4.10}
\end{equation}	
Moreover, when a Poisson approximation is used, the approximated heritability takes the form given in (\ref{Eq.4.7}).

We turn now to the calculation of the heritability of the accumulated hazard function evaluated at a time point $t$. In the continuous time case we define the accumulated hazard function, for each $t\in \rep$, by
\begin{equation}
\nonumber
\Lambda (t) = \int_0^t Y(s) \lambda (s) ds
\, ,
\end{equation}
 and in the discrete time case  the accumulated hazard function is given, for each $t\in \Zp$, by
\begin{equation}
\nonumber
\Lambda (t) = \sum_{s\leq t} Y(s) \lambda (s) 
\, ,
\end{equation}
where $\lambda ( \cdot ) $ is the hazard function in play.
Furthermore, the conditional random variable
\begin{equation} \label{Eq.4.13B}
{\cal{Z}}(t) = N(t)  \, \vert \, T \geq t, X(t) 
\end{equation}
is such that $E\left ({\cal{Z}}(t)\, \vert \, \bU = \bu , \, \bV = \bv \right ) = \Lambda \left (t \,\vert \, \bU = \bu , \, \bV = \bv \right )$. Therefore,  ${\cal{Z}}(t)$ will be treated as the trait of interest, for which we calculate the phenotypic variance and the heritability. A more precise definition of ${\cal{Z}}(t)$ using counting processes theory is given in appendix \ref{ref.appendix.01}. It can be shown (see appendix \ref{ref.appendix.01}) that in the case of the continuous time the variance of ${\cal{Z}}(\cdot )$ evaluated at a given $t\in\rep$ is approximated by
\begin{equation}
Var \left[ {\cal{Z}}(t) \right] \approx \left[ \Lambda^*(t) \right]^2 \left( \sigma_g^2 + \sigma_e^2 \right) + \phi \Lambda^*(t)
\label{Eq.4.13}
\end{equation}
where $\Lambda^*(t) = \int_{0}^{t} Y(s) \lambda^*(s)ds$. Therefore, the heritability is approximated by 
\begin{equation}
h_{ \Lambda(t) }^2 
\approx  \frac{ \sigma_g^2 }{ \sigma_g^2 + \sigma_e^2 + \phi \left[ \Lambda^*(t) \right]^{-1} }
\,\, .
\label{Eq.4.14}
\end{equation}	

In the case of the discrete time the variance of ${\cal{Z}}(\cdot )$ evaluated at a given $t\in\Zp$ is approximated by 
\begin{equation}
Var \left[ {\cal{Z}}(t) \right] \approx \left[ \Lambda^*(t) \right]^2 \left( \sigma_g^2 + \sigma_e^2 \right) + \phi \gamma(t),
\label{Eq.4.15}
\end{equation}
where $\gamma(t) = \sum_{i=0}^{t}{\phi Y(t)\lambda^*(t)[1 - \lambda^*(t)]}$  and the cumulative hazard is $\Lambda^*(t) = \sum_{s=0}^{t} Y(s){\lambda^*(s)}$ .Then, the heritability is approximated by
\begin{equation}
h_{ \Lambda(t) }^2 
\approx  \frac{ \sigma_g^2 }
         { \sigma_g^2 + \sigma_e^2 +\phi \gamma(t) [\Lambda^*(t)]^{-2}}
         \,\, .
\label{Eq.4.16}
\end{equation}	
When the Poisson approximation is used the heritability is approximated by the right side of (\ref{Eq.4.14}).

The calculations above are performed for one individual. We propose that given a population of $n$ individuals, the phenotypic variances and the heritabilities for the hazard function or the cumulative hazard (evaluated at a time $t$) should be calculated using the estimated variance components and an estimated hazard or an estimated cumulative hazard as described below. In the continuous time case, where the PCHM is used, the observed time is split in intervals $I_k = [ t_k, t_{k + 1})$, for $k = 0,\dots,K$  where $0 = t_0 < t_1 < \dots < t_K  < t_{K+1} = \tau$, where $\tau$ is the largest observed time.  Given that $\bU = \hat{\bu}$ and $\bV = \hat{\bv}$, where $\hat{\bu}$ and $\hat{\bv}$ are the BLUP of the random components, the conditional hazard is defined, for each $t \in I_k$, by 
\begin{eqnarray} \nonumber
\lambda_i^*(t) = \lambda_k \exp \left[ \bX_{i,k}^{'}\bbeta + \bZ_{i}^{'}\hat{\bu} + \bW_{i}^{'}\hat{\bv}\right] = \exp \left[ \eta_{i,k}(t) \right]
\,\, ,
\end{eqnarray}
where $\eta_{i,k}(t) =  \log(\lambda_k) +  \bX_{i,k}^{'}\bbeta + \bZ_{i}^{'}\hat{\bu} + \bW_{i}^{'}\hat{\bv}$ and $\bX_{i,k} = \bX_i(t)$ for $t \in I_k$. We obtain $\hat{\eta}_{i,k}(t) = \widehat{\log(\lambda_k)} + \bX_{i,k}^{'} \hat{\bbeta} + \bZ_{i}^{'}\hat{\bu} + \bW_{i}^{'}\hat{\bv} $ from the adjusted model and then calculate, for each time $t \in I_k$,
\begin{eqnarray} \nonumber
\tilde{\lambda}^*(t) = \tilde{\lambda}^*_k = \exp \left[ \bar{\eta}_k \right]
\,\, ,
\end{eqnarray}
where $ \bar{\eta}_k = n^{-1}\sum_{i=1}^{n} {\hat{\eta}_{i,k}}$. The cumulative hazard is estimated, for $t \in I_k$, by
\begin{eqnarray} \nonumber
\tilde{\Lambda}^*(t) = \sum_{j= 0}^{k - 1} \Delta_j \tilde{\lambda}_j^* + (t - t_k)\tilde{\lambda}_k^* \,\, ,
\end{eqnarray}
where $\Delta_j = t_{j+1} - t_j$ is the length of the interval $I_j$.

In the discrete time case the hazard, conditionally that $\bU = \hat{\bu}$ and $\bV = \hat{\bv}$, is  given, for each $t \in \mathbb{Z}_+$, by 
\begin{eqnarray} \nonumber
 \lambda_i^*(t) = \lambda_t \exp \left[ \bX_i^{'}(t)\bbeta + \bZ_{i}^{'}\hat{\bu} + \bW_{i}^{'}\hat{\bv} \right] = \exp \left[ \eta_i(t) \right]
 \,\, ,
\end{eqnarray}
where $\eta_i(t)  = \log(\lambda_t) +  \bX_i^{'}(t)\bbeta + \bZ_{i}^{'}\hat{\bu} + \bW_{i}^{'}\hat{\bv}$.
We obtain $\hat{\eta}_i(t) = \widehat{ \log(\lambda_t)} + \bX_i^{'}(t)\hat{\bbeta} + \bZ_{i}^{'}\hat{\bu} + \bW_{i}^{'}\hat{\bv}$ from the adjusted model and then calculate for each time $t$ ($t = 0, 1,2,\dots, $)
\begin{eqnarray} \nonumber
\tilde{\lambda}(t) = \exp \left[ \bar{\eta}(t) \right] \,\, ,
\end{eqnarray}
where $ \bar{\eta}(t) = n^{-1}\sum_{i=1}^{n} {\hat{\eta}_i(t)}$. The cumulative hazard is estimated by $\tilde{\Lambda}^*(t ) = \sum_{j = 0}^{t} \tilde{\lambda}^*(j)$. 
The heritabilities will be estimated at $\tilde{\lambda}^*(t_m)$ and $\tilde{\Lambda}^*(t_m)$ where $t_m$ represents the median survival time, {\it i.e.} $P(T \geq t_m) = 0.5$ in the illustrative examples.
  
%%%%%%%%%%%%%%%%%%%%%%%%%%%%%%%%%%%%%
\section{Two illustrative examples}
\label{Sect.4}
%%%%%%%%%%%%%%%%%%%%%%%%%%%%%%%%%%%%%

The methods presented above will be illustrated by two examples: one involving the longevity of sows (section \ref{Sect.4.1}) and another characterizing the longevity of dairy cattle (section \ref{Sect.4.2}). In both examples,  we will characterize the longevity of female animals in two different ways: first, by the length of the productive life of the animals (sows or cows) measured by the number of days from the first parity to the culling day (ND); second, by the number of survived parities (NP). Models based on continuous time and models based on discrete time were applied to study ND and NP, respectively.

Incomplete observations are present in both examples because some animals were still alive at the end of the study, some animals were moved, sold or exported during the study and due to intentional right truncation. In the study of sow longevity the proportion of censure was relatively low ($9.6\%$), while in the study of longevity of cows a much higher proportion of censure was observed ($23.5\%$). Furthermore, in the study of longevity of cows it was possible to distinguish two culling causes: slaughter ($67.7 \%$ of the observations) and death (only $8.8\%$ of the observations), which characterizes a typical scenario of competing risks. Therefore, it is natural to use bivariate models in this example to model these two culling causes simultaneously. Bivariate models will be used in the example of sows longevity to model ND and NP simultaneously and use that to discuss how much the information of these two characterizations of the longevity overlap.

%------------------------------------------------------------------------------------------------
\subsection{Longevity of sows}
\label{Sect.4.1}
%------------------------------------------------------------------------------------------------

The data in this example was retrieved from the registers of the Danish Pig Center of 200,084 pure Landrace sows that were giving birth in the years between 1999 and 2010.  Only the sows that had the first parity in the comprised period were included in the study. All the models proposed for ND and NP included the following explanatory variables: age at the first parity, year and season at the first parity, litter size (total number of piglets born per parity) and herd year size (total number of sows farrowing per herd per year). The litter size and the herd year size were both time dependent variables. In addition two random components were included: a sire component with pedigree, representing the sire additive genetic effects and a herd-year component representing the environment effects. Since a sire model was used, the additive genetic variance was estimated by multiplying the additive sire variance by four. 

The Kaplan-Meyer estimate of the median survived NP was 2 parities (CI95\% 2;3) and of the median survived ND was 320 days (95\% CI, 319 - 322). The median number of days between two consecutive parities for the sows that did not die in the respective period varied from 152 to 155 days. Thus, the median survived time of 320 days from the first parity is approximately the time necessary for the sow, give it had one parity, have the second and survive until a third parity.

\begin{figure}[h]
	\begin{center}
		\resizebox*{8.9cm}{!}{\includegraphics{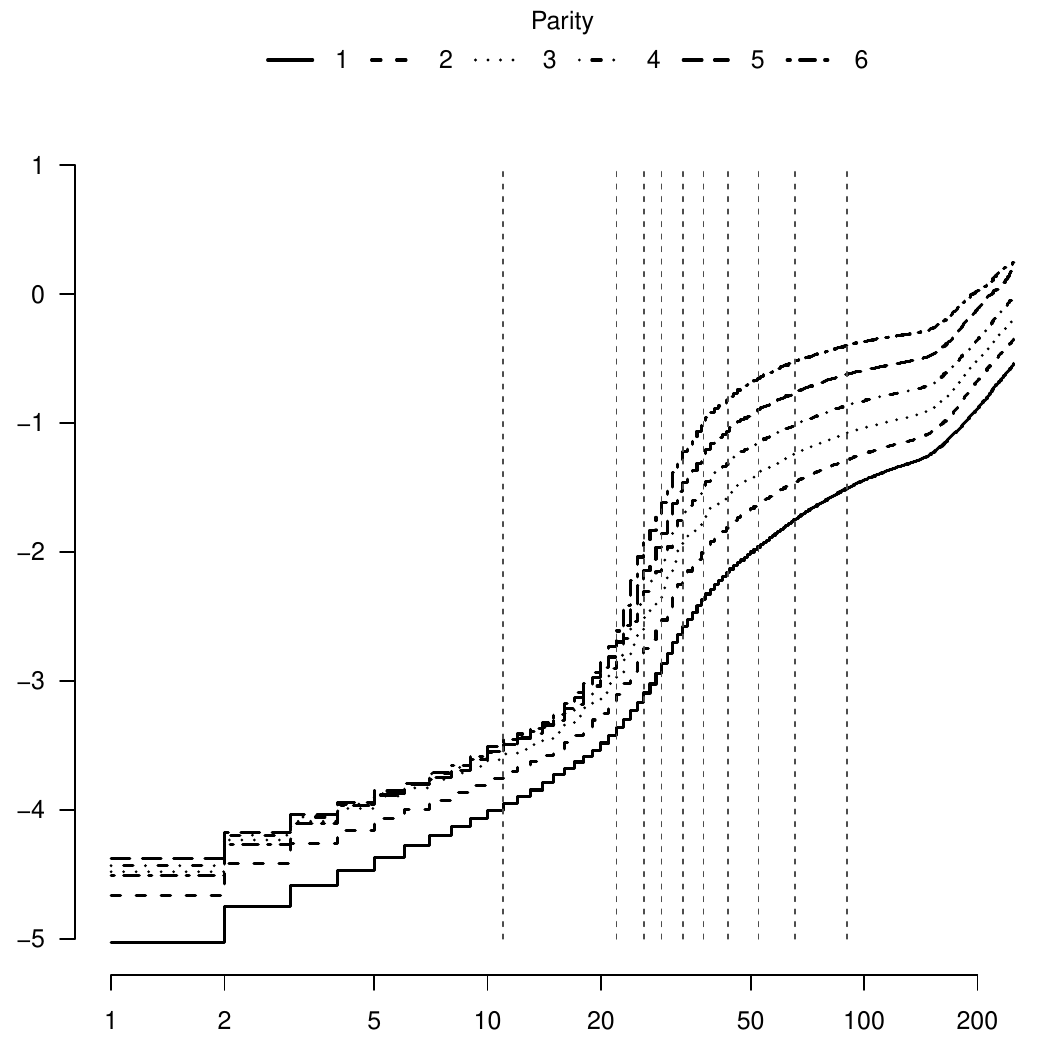}}
		\caption{Log of cumulative hazard curves for the number of days trait(ND) stratified by parity at the sow's longevity study. Axis x is presented on the logarithm of days scale. The vertical dashed lines are the cut points for the intervals where the baseline function was assumed to be constant.}
	\label{fig.5.1}
	\end{center}
\end{figure}

Figure \ref{fig.5.1} displays the logarithm of the cumulative hazard curves for ND stratified by parity. The curves per parity presented similar behavior, however, they are not parallels indicating a non proportional effect of parity. The dashed lines represents the cut-points of the intervals where the baseline hazard function was assumed to be constant.

Table \ref{tab.5.1} presents the estimates of the variance components and heritabilities from the piece-wise constant hazard model stratified by parity based on the ND and for two discrete relative risk models based on the NP (an exact model and an approximated model via Poisson approximation). The estimates of variance for the sire and the herd-year components were very similar among the three adjusted models ($\sim$ 0.065 with SE $\sim$ 0.003 for the sire component and $\sim$ 0.220 with SE $\sim$ 0.009 for the herd-year component). Comparing the execution time between the two adjusted discrete models the approximated models was around 72 times faster compared to the time spent to execute the exact model.

A bivariate model  describing the ND and the NP traits (using a Poisson approximation for the discrete marginal models) was fitted (see Table \ref{tab.5.2}). 
%##### Table 1 #############################################################################
\begin{table}[!ht]
  \caption{Estimated variance components (with asymptotic standard error in parenthesis) and heritabilities for the number of days (ND) and for the number of parities (NP) at the sows longevity study.}
  { 
		\begin{tabular}{lrrr}
    \toprule[1.5pt]
    \bf Source       				 &\bf ND-PCHM$^{\rm a}$&\bf NP-DRRM$^{\rm b}$&\bf NP-DRRM$_{P}^{\rm c}$\\
		\midrule
    \bf Sire                 & 0.071 (0.005)       & 0.066 (0.002) 			 & 0.063 (0.003) \\
    \bf Herd-Year            & 0.217 (0.010)       & 0.220 (0.005)       & 0.219 (0.009) \\
    \bf Dispersion           & 3.089 (0.002)       & 0.978 (0.002)       & 0.653 (0.001) \\
		\midrule
    $h_{\lambda}^2$$^{\rm d}$& -                   & 0.098               & 0.096         \\
    $h_{\Lambda}^2$$^{\rm e}$& 0.180               & 0.161               & 0.165         \\
		\midrule
    \bf Execution time$^{\rm e}$& 48.5 min         & 8h 28 min           & 7 min         \\
		\bottomrule[1.5pt] 
		\end{tabular}
	}
	
	\begin{flushleft}
	$^{\rm a}$ PCHM: Piece-wise constant hazard model stratified by parity. \\
	$^{\rm b}$ DRRM: Discrete relative risk model. \\
	$^{\rm c}$ DRRM$_{P}$: Discrete relative risk model via Poisson approximation. \\
	$^{\rm d}$ Estimated heritability for the hazard ($\lambda$) at the median survival time. \\
	$^{\rm e}$ Estimated heritability for the cumulative hazard ($\Lambda$) at the median survival time. \\
	$^{\rm f}$ Execution time (Intel i5 processor).
\end{flushleft}

  \label{tab.5.1}
\end{table}
%###########################################################################################

%##### Table 2 #############################################################################
\begin{table}[!ht]
  \caption{Estimated variance components and correlations from the bivariate model based on the number of days (ND) and number of parity (NP) at the sows longevity study.}
  { 
		\begin{tabular}{lrrrrr}
    \toprule[1.5pt]
    \bf Source     & \bf Trait & $\sigma^2$ (\bf SE) & \bf Cor (\bf SE)  \\
    \midrule
    \bf Sire 	     & \bf ND    & 0.022 (0.001)       & \multirow{2}{*}{1.000 (0.004)} \\
                   & \bf NP    & 0.081 (0.004)       &                                \\
									 &	         &                     &                                \\
    \bf Herd-Year  & \bf ND  	 & 0.332 (0.015)       & \multirow{2}{*}{1.000 (0.002)} \\
                   & \bf NP  	 & 0.231 (0.010)       &                                \\
									 &           &                     &                                \\
    \bf Dispersion & \bf ND    & 3.130 (0.002)       & \multirow{2}{*}{-}             \\
                   & \bf NP    & 0.657 (0.001)       &                                \\
		\bottomrule[1.5pt]
    \end{tabular}
	}	
  \label{tab.5.2}
\end{table}
%###########################################################################################

%##### Figure 1 #############################################################################
%###########################################################################################

%###########################################################################################

\clearpage

%------------------------------------------------------------------------------------------------
\subsection{Longevity of dairy cattle}
\label{Sect.4.2}
%------------------------------------------------------------------------------------------------

The data analyzed in this example was retrieved from the register of Knowledge Centre for Agriculture of 142,133 Jersey Danish dairy cows that were calving in the period between 1990 to 2006. The explanatory variables (fixed effects) included in the models were: herd year size (number of calving per herd per year), year and season of each parity, as time dependent variables and age at the first parity and a coefficient of general heterosis (defined as a continuous explanatory variables), as time independent variables. Two random components were included in the model, a combination of herd-year-season component representing the environment effects and a sire component with the pedigree representing the genetic effect. In addition, the cows could be culled by one of two possible reasons: death and slaughter. 

\begin{figure}[h]
	\begin{center}
		\resizebox*{14cm}{!}{\includegraphics{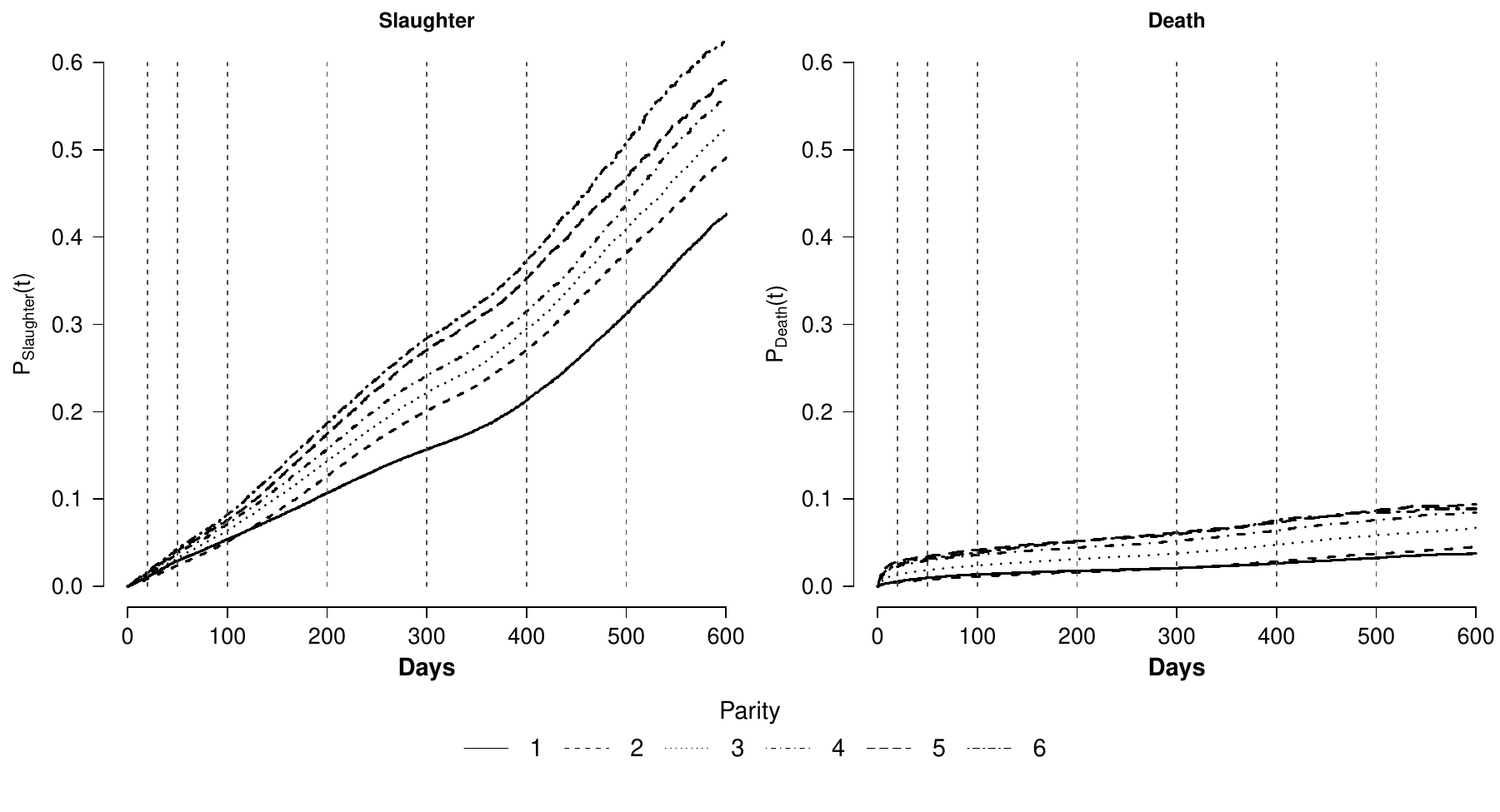}}
		\caption{Cumulative incidence probability curves for the ND stratified by parity for the cow's longevity study. The dashed vertical lines are cut points for the intervals were the hazard was assumed constant. $P_J(t) = P(T \leq t, J = j)$ for $J = (Death, Slaughter)$.}
	\label{fig.5.2}
	\end{center}
\end{figure}

The Kaplan-Meyer estimate of the overall median NP was 3 parities (95\% CI, 3;3) and overall median ND was 826 days (95\% CI,  820 - 831). The median number of days between 2 consecutive parities for the cows that did not die in the respective period varied from 361 days to 367 days.  Figure \ref{fig.5.2} displays the stratified cumulative incidence probability curves for ND for the slaughter and the death, respectively. The dashed lines represent the cut points for the intervals where the baseline hazard were assumed to be constant for both specific reasons.

Tables \ref{tab.5.3} displays the estimates of the variance and covariance components from the competing risk models describing the ND and NP traits. The sire end herd-year-season variances for death rate based on the ND trait were very small, both for the death and the slaughter rate. In contrast, the discrete time model detected significantly larger variances of the random components. The failure of the continuous time models can be explained by the large amount of censure in the data.

%##### Table 3 #############################################################################

\begin{table}[!ht]
  \caption{Estimated variances and covariances components for the number of parities (NP) and the number of days (ND) from a discrete and continuous multivariate competing risk model, respectively, at the dairy cattle longevity study.}
  { \footnotesize 
		\begin{tabular}{lrrrrr}
		\toprule[1.5pt]
		                 &             &\mc{\bf NP}                            &\mc{\bf ND}                            \\
		\bf Source       &\bf Cause    &$\sigma^2$ (\bf SE)&Cov (\bf SE)       &$\sigma^2$ (\bf SE)&Cov (\bf SE)       \\
		\midrule
		\bf Sire         &\bf Death    & 0.102 (0.041)     &\-0.002 & 0.7$\times 10^{-3}$ (0.006)   & 0.002 \\
                     &\bf Slaughter & 0.029 (0.003)     &   (0.004)              & 0.008 (0.002)   & (0.003)                    \\
		 & & & & & \\																												
    \bf HYS $^{\rm a}$&\bf Death    & 0.458 (0.015)     &-0.055  & 0.1$\times 10^{-5}$ (0.079)   & 0.7$\times 10^{-7}$ \\
                     &\bf Slaughter& 0.061 (0.002)     &      (0.004)             & 0.3$\times 10^{-6}$ (0.010)   & (0.021)              			 \\
		 & & & & & \\																		
    $\phi_j$ $^{\rm b}$  &\bf Death    & 0.691 (0.002)     &\mr{-}             & 5.165 (0.005)   &\mr{-}               \\
										 &\bf Slaughter& 0.710 (0.002)     &                   & 4.148 (0.004)   &                     \\
    \midrule
		\bf $h_{\lambda}^2$$^{\rm c}$&\bf Death      &\mc{0.013} &\mc{-}\\
		                             &\bf Slaughter &\mc{0.040} &\mc{-}\\
		 & & & & & \\	
		\bf $h_{\Lambda}^2$$^{\rm d}$&\bf Death      &\mc{0.044} &\mc{0.2$\times 10^{-3}$}\\
		                             &\bf Slaughter  &\mc{0.108} &\mc{0.012}\\	 
		\bottomrule[1.5pt]
    \end{tabular}
  }	
	\begin{flushleft}
	$^{\rm a}$ Herd-year-season random component.\\
	$^{\rm b}$ Marginal dispersion parameters.\\
	$^{\rm c}$ Marginal heritabilities for the hazard ($\lambda$) at the median overall survival time.\\
	$^{\rm d}$ Marginal heritabilities for the cumulative hazard ($\Lambda$) at the median overall survival time.
	\label{tab.5.3}
\end{flushleft}

\end{table}

%##### Figure 4 #############################################################################
%##### Figure 4 #############################################################################

\clearpage

%%%%%%%%%%%%%%%%%%%%%%%%%%%%%%%%%%%%%
\newpage
\section{Discussion}
\label{Sect.5}
%%%%%%%%%%%%%%%%%%%%%%%%%%%%%%%%%%%%%

The several variants of models for studying longevity in quantitative genetics can be described using the following general approach. The models are one-dimensional and describe the conditional hazard function, given the random components, as taking the form
\begin{eqnarray} \nonumber
	    \lambda_{i}(t \vert \bU=\!\bu , \! \bV\!=\!\bv)
	     = 	
	    \lambda_{0}(t) \exp \left ( \bX_{i}^{'}(t)\bbeta + \bZ_{i}^{'}\bu + \bW_{i}^{'}\bv \right )
	 \,  ,
\end{eqnarray}	
for $t$ in $\rep$ or in $\Zp$.
The models based on continuous time used in the literature differ essentially in the form of the baseline hazard function $\lambda_{0}(\cdot )$: the classical Cox model (with frailties) assumes the hazard function to be a smooth function (almost everywhere) imposing in this way almost no restriction on $\lambda_{0}(\cdot)$ (see \cite{kalb02, pankratz00, ripatti00});  the piece-wise Weibull models (see \cite{ducroq88}) assume  $\lambda_{0}(\cdot)$ to be a saw function (\ie a piecewise linear function); while the piece-wise constant hazard models (as described here) assume $\lambda_{0}(\cdot)$ to be a step function (\ie a piece-wise constant function). These models are nested, in the sense that saw functions are smooth almost everywhere, and step functions are particular cases of saw functions. Moreover, smooth functions can be arbitrarily approximated  (point-wisely) by saw functions or by step functions. A fourth category of models, which was introduced here, are the piece-wise constant hazard models with free dispersion parameter. These models are an instance of quasi-likelihood based models (see \cite{wedd74, bresl93}). They contain the traditional piece-wise constant hazard models as a particular case (by setting the dispersion parameter $\phi$ to be constant equal to 1) and might arbitrarily approximate the piece-wise Weibull models and the Cox frailty models. Although in many practical situations those models yield equivalent representations of the data and produce similar results, this is not always the case for the quantitative genetics applications discussed here. 

One of the interesting features of the piece-wise constant hazard model with (free) dispersion parameter presented here is that the phenotypic variance can be approximately decomposed as the sum of parcels representing the additive genetic effects, environmental effects and unspecified sources of variability. This is analogous to the classic decomposition of the phenotypic variance obtained in gaussian linear mixed genetic models. Here, the component of the phenotypic variance involving the dispersion parameter $\phi$ plays the rule of the residual variance in the classic models. Note that, if we fix the dispersion parameter $\phi$, say by setting it equal to 1 as in the piece-wise constant hazard model, it might be that we obtain models that describe the data well and the referred decomposition phenotypic variance would still exists, but this would be equivalent to set the residual variance to be constant, which clearly causes problems in the representation of quantitative genetic phenomena. For instance, in a sire model the genetic variance transmitted by the dam is not captured by the random component representing the sire, therefore, this part of the variability should be represented in the residual variance (\ie should compose part of the residual variance); but this is not possible to occur if the residual variance is set to be constant. This issue occurs also in the piece-wise  Weibull models (see  \cite{ducroq88}). Furthermore, in the most general case where $\lambda_{0}$ is only assumed to be smooth (the Cox frailty model) it might be argued that $\lambda_{0}$ could be point wisely approximated (almost every where) by a sequence of step functions obtained even from distributions in a model with dispersion parameter, which would have a proper decomposition of the phenotypic variance.  However, in this case it would not be possible to identify the dispersion parameter $\phi$ (as essentially argued in \cite{giolo11}); therefore, one would still have the representation issues referred above.

The piece-wise constant hazard model and the piece-wise Weibull model are parametric models, making it possible to implement relatively efficient inference techniques.  This explains why they are implementable for complex and large models as in quantitative genetics typical applications.  Moreover, when the baseline hazard function $\lambda_{0} (\cdot )$ is assumed to be piece-wise linear (a saw function), as in the piece-wise Weilbull models, the rate of approximation of $\lambda_{0} (\cdot )$  to a smooth function (as in the frailty model) is improved (specially if there are regions where the smooth baseline is steep) as well argued in \cite{ducroq88}; however, in this case the inference via generalized linear mixed models would not be possible without using profile likelihood techniques (to estimate the shape parameter), which rules out the direct use of  extensions containing a dispersion parameter or multivariate extensions already implemented. Finally, a disadvantage of the piece-wise constant hazard models and the piece-wise Weibull models is that both depend on an arbitrary choice of time cutting points.

Models based on discrete time (\ie variants of the discrete relative risk models)  has been occasionally used in quantitative genetic previously; however, without incorporating a dispersion parameter. These models do not assume any pre-specified form for the baseline hazard function and also not require any arbitrary definition of time cutting points; therefore, they are similar to the Cox frailty models. Due to the coincidence with the Bernoulli models with random components (with or without dispersion parameter), techniques for parameter inference are at hand, allowing for efficient implementations capable to handle complex large problems as illustrated here. Moreover, these models allow to incorporate time-dependent variables (by splitting data with record for each survived time for each individual) both as fixed effects and as random components. The Poisson approximation presented here might represent a substantial save in computational resources and yields essentially the same results as the exact inference, as illustrated in the example of sows longevity presented in section \ref{Sect.4.1}.

The continuous time models and the discrete time models yielded similar results in the example of sows longevity. However, the continuous time models failed in the example of dairy cattle longevity (section \ref{Sect.4.2}), which can be explained by the fact that, in this case, both rare events and high frequencies of censored observations are observed. This is in line with the limitations of the continuous time models discussed in section \ref{Sect.2.4} and shows that those limitations do occur in practical situations.

The lack of multivariate techniques for analyzing several traits simultaneously has been a serious limitation in the application of survival models in quantitative genetics \cite{sere06, ducroq99}. This lack not only makes the study of the relation between longevity and other traits impossible, but also limits very much the possibility of performing time-to-event-analysis in the presence of competing risks. Indeed, if there are two competing culling reasons, one might use standard survival models for modeling the rate of occurrence of one of the culling reasons by considering the observations where the other culling reason occurred as censored. However, this naive approach implies in the implicit use of the assumption that the competing culling reasons are independent. The example of dairy cattle longevity presented here illustrates that this assumption might be not reasonable. Indeed, in this example we detected a significant negative correlation between the herd-year-season random components; moreover, a range of fixed effects were found to be significant for both culling reasons, implying that a naive use of marginal models would violate the assumption of absence of informative censoring.  

%%%%%%%%%%%%%%%%%%%%%%%%%%%%%%%%%%%%%%%%
\section*{Acknowledgements}
%%%%%%%%%%%%%%%%%%%%%%%%%%%%%%%%%%%%%%%%

Rafael Pimentel Maia was financed by the project  "Svineavl: Developing New Methods for Genetic Selection of Sow Durability" , Ministry of Food, Agriculture and Fisheries of Denmark.

%%%%%%%%%%%%%%%%%%%%%%%%%%%%%%%%%%%%%%%%

\newpage
%%%%%%%%%%%%%%%%%%%%%%%%%%%%%%%%%%%%%%%%
%%%%%%%%%%%%%%%%%%%%%%%%%%%%%%%%%%%%%%%%
\appendix
%%%%%%%%%%%%%%%%%%%%%%%%%%%%%%%%%%%%%%%%

%%%%%%%%%%%%%%%%%%%%%%%%%%%%%%%%%%%%%%%%
\section{Technical details on the decomposition of the phenotypic variance}
\label{ref.appendix.01}
%%%%%%%%%%%%%%%%%%%%%%%%%%%%%%%%%%%%%%%%

Let $T$ be a non-negative random variable representing the observed survival time and $D$ be an indicator variable taking the value 1 if a death is observed and 0 otherwise (censure). Denote by $\bX(t) = \{ X(s): 0 \leq s < t \}$ the trajectory function of the known vector of explanatory variables until the time just before $t$. Additionally, we consider two independent gaussian random components, $\bU$ and $\bV$, representing the additive genetic effects and some environmental effects, respectively. 

We define the death counting process $\mathbf{N} = \left \{ N(t) : t \in \rep \mbox{ or } \Zp \right \}$, where, for each time $t$, $N(t) = \mathbf{1}(T \leq t, D = 1)$. The increment of the death process at the time $t$ is given by $dN(t) = N(t) - N(t-)$, where $N(t-)= \lim_{ \Delta \downarrow 0} N(t - \Delta)$. The at risk process is given by $\mathbf{Y} = \left \{ Y(t) : t \in \rep \mbox{ or } \Zp \right \}$, where, for each time $t$, the random variable $Y(t)$ takes the value $1$ when the individual is at risk at time $t$ or $0$ otherwise.  

The death counting process and the at risk process are both adapted to the filtration $\mathbf{F} = \left \{ {\cal{F}}_{t^-} : t \in \rep \mbox{ or } \Zp \right \}$, where, for each time $t$, ${\cal{F}}_{t^-} = \sigma \left\{ N(s), Y(s), X(s), 0 < s < t \right\}$ is the $\sigma$-algebra generated by $N(\cdot ), Y(\cdot ), X(\cdot )$ up to $t-$ (\ie up to the time just before $t$) . Here, ${\cal{F}}_{t^-}$ represents what is known up to (but not including) time $t$.
% (for more details see \cite{kalb02} page 151).

Conditional on the random components $\bU$ and $\bV$, under the standard independent censoring assumption (see \cite{andersen93, kalb02}), we have that
\begin{equation}
P \left[ dN(t) = 1 \vert \bU = \bu, \bV = \bv, {\cal{F}}_{t^-} \right] 
= Y(t)\lambda (t \vert \bu, \bv),
\label{Eq.A.2}
\end{equation}
where $\lambda(t \vert \bu, \bv) = \lambda_0(t) \exp \left[ X(t)^{'}\beta + Z\bu + W\bv \right]$ and $Y(t)\lambda (t \vert \bu, \bv)$ is the intensity associated to the counting process $\mathbf{N}$ at time $t$. Moreover, the cumulative intensity of $\mathbf{N}$ up to time $t$ is $\Lambda(t \vert \bu, \bv) = \int_{0}^{t}{Y(s)\lambda(s \vert \bu, \bv)} ds$. Note that 
$\mathbf{\Lambda} = \{ \Lambda(t \vert \bu, \bv) : t \in \rep \mbox{ or } \Zp  \}$ is predictable with respect to the filtration  $\mathbf{F}$ . Define, for each time $t$, 
\begin{equation}
M(t) = N(t) - \Lambda(t) 
\,\, .
\label{Eq.A.3}
\end{equation}
The process $\mathbf{M}= \{M(t) : t \in \rep \mbox{ or } \Zp \}$ is a martingale. Here, (\ref{Eq.A.3}) is the Doob-Meyer decomposition of the death counting process where $\mathbf{\Lambda}$ is the compensator. Define, for each time $t$, the increment of the martingale   $\mathbf{M}$ by $dM(t) = M(t)- M(t-)$, where $M(t-)= \lim_{ \Delta \downarrow 0} M(t - \Delta)$. Since $\mathbf{M}$ is a martingale, $E[dM(t) \vert {\cal{F}}_{t^-}]  = 0$ for each time $t$, which shows that $\mathbf{M}$ plays the role of the residuals of the model. The predictable variation process of the martingale $\mathbf{M}$ at the time $t$ is
\begin{eqnarray} \nonumber
\left\langle M \right\rangle (t) = \int_{0}^{t}{Var [ dM(s) \vert {\cal{F}}_{s^-} ]}
\end{eqnarray}
and
\begin{eqnarray} \nonumber
d\left\langle M \right\rangle (t) = Var [ dM(t) \vert {\cal{F}}_{t^-} ]
\,\, .
\end{eqnarray}

Now, we study the decomposition of the phenotypic variance related to the hazard function evaluated at a given time $t$. Define the conditional random variable 
\begin{equation} \label{eq.A.4}
{\cal{Y}}(t) = dN(t) \vert {\cal{F}}_{t^-}
\,\, ,
\end{equation}
which is equivalent to ${\cal{Y}}(t)$ informally defined by (\ref{eq.4.1}) in section \ref{Sect.3}. 

 Given the random components $\bU$ and $\bV$, 
 \begin{equation}
 E \left[ {\cal{Y}}(t) \vert \bU, \bV \right] = Y(t)\lambda(t\vert \bu, \bv)
 \label{Eq.A.5}
\end{equation}
 and
\begin{equation}
Var \left[ {\cal{Y}}(t) \vert \bU, \bV \right] =  \phi Var \left[ dM(t) \vert \bU, \bV, {\cal{F}}_{t^-} \right],
\label{Eq.A.6}
\end{equation}
where $\phi$ is the dispersion parameter. Using (\ref{Eq.A.5}) and  (\ref{Eq.A.6}), the variance of ${\cal{Y}}(t)$ is given by
\begin{eqnarray}
 \label{Eq.A.7}
Var[ {\cal{Y}}(t) ] 
&=& Var \left\{ E \left[ {\cal{Y}}(t) \vert \bU, \bV \right] \right\} + E \left\{ Var \left[ {\cal{Y}}(t) \vert \bU, \bV \right] \right\}  \\ \nonumber
&=& Var \left\{ Y(t)\lambda(t\vert \bU, \bV) \right\} + E \left\{ \phi Var [ dM(t) \vert \bU, \bV, {\cal{F}}_{t^-} ]\right\}
\,\, .
\end{eqnarray}
A Taylor expansion argument (see \cite{hald1952}, page 118) yields  
\begin{eqnarray} \nonumber
 E \left\{ Var\left [ {\cal{Y}}(t) \vert \bU, \bV \right] \right \}
  & \approx & 
   Var \left [ Y(t) \, \vert \, \bU={\bf 0}, \bV={\bf 0} \right ]  \\
  & =  &\phi Var \left[ dM(t) \vert \bU ={\bf 0}, \bV ={\bf 0} , {\cal{F}}_{t^-} \right]
  \,\, .
  \label{Eq.A.8}
\end{eqnarray}
Moreover, a second order Taylor expansion (see \cite{hald1952}, page 118) yields
\begin{eqnarray} \nonumber
 Var \left\{ E \left[ {\cal{Y}}(t) \vert \bU, \bV \right] \right\}
  & \approx & 
  \sigma_g^2
  \left \{    
     \frac{\partial}
            {\partial \bu}
     E \left[ {\cal{Y}}(t) \vert \bU, \bV \right]
  \right \}_{\left ( \bU={\bf 0}, \bV={\bf 0}\right )}^2
  \\ \nonumber
  & & +
  \sigma_e^2
    \left \{    
     \frac{\partial}
            {\partial \bv}
     E \left[ {\cal{Y}}(t) \vert \bU, \bV \right]
  \right \}_{\left (\bU={\bf 0}, \bV={\bf 0}\right )}^2
  \\ &  &  \nonumber \\ & = &
   \left[ Y(t) \lambda^*(t) \right]^2 \left( \sigma_g^2 + \sigma_e^2 \right)
  \,\, ,
  \label{Eq.A.9}
\end{eqnarray}
where $\bu$ and $\bv$ are realisations of the random components. Moreover, $ \lambda^*(t) = \lambda(t \vert \bU = \bf 0, \bV = \bf 0)$.

In the continuous time case, for each $t \in\rep$, 
\begin{eqnarray} \nonumber
Var \left[ dM(t) \vert \bU ={\bf 0}, \bV ={\bf 0} , {\cal{F}}_{t^-} \right] = Y(t)\lambda^*(t)
\,\, , 
\end{eqnarray}
(see \cite{kalb02, andersen93}) and therefore, using (\ref{Eq.A.8}) and (\ref{Eq.A.9}) yields
\begin{eqnarray}
\nonumber
Var \left[ {\cal{Y}}(t) \right] \approx Y(t) \left\{ \left[  \lambda^*(t) \right]^2 \left( \sigma_g^2 + \sigma_e^2 \right) + \phi \lambda^*(t) \right\}
\,\, ,
\end{eqnarray}
which proves the decomposition of the phenotypic variance given in (\ref{Eq.4.6}).

In the discrete time case, for each $t \in\Zp$, 
\begin{eqnarray} \nonumber
Var \left[ dM(t) \vert \bU ={\bf 0}, \bV ={\bf 0} , {\cal{F}}_{t^-} \right] 
= 
Y(t)
\lambda^*(t) \left [ 1- \lambda^*(t) \right ]
\,\, , 
\end{eqnarray}
(see \cite{kalb02, andersen93}) and therefore, using (\ref{Eq.A.8}) and (\ref{Eq.A.9}) yields
\begin{eqnarray}
Var \left[ {\cal{Y}}(t) \right] \approx Y(t) \left\{ \left[  \lambda^*(t) \right]^2 \left( \sigma_g^2 + \sigma_e^2 \right) + \phi \lambda^*(t)  \left [ 1- \lambda^*(t) \right ] \right\}
\,\, ,
\label{Eq.A.10}
\end{eqnarray}
which proves the decomposition of the phenotypic variance given in (\ref{Eq.4.9}).

For the decomposition of the phenotypic variance related to the cumulative hazard function evaluated at a given time $t$ we define the conditional random variable 
\begin{eqnarray} \nonumber
{\cal{Z}}(t) = N(t) \vert {\cal{F}}_{t^-}
\,\, .
\end{eqnarray}
This is equivalent to conditional random variable ${\cal{Z}}(t)$ informally defined in (\ref{Eq.4.13B}). Note that,
\begin{eqnarray} \nonumber
E \left[ {\cal{Z}}(t) \vert \bU, \bV \right] = \Lambda( t \vert \bU, \bV)
\mbox{ and }
 Var \left[ {\cal{Z}}(t) \vert \bU, \bV \right] = \langle M \rangle (t) \vert \bU, \bV
 \,\, .
\end{eqnarray}
The decomposition of the variance of ${\cal{Z}}(t)$ is obtained by applying an analogous calculation as in the decomposition of the variance of ${\cal{Y}}(t)$. This yields the equations (\ref{Eq.4.13}) and (\ref{Eq.4.15}) for the decomposition of the phenotypic variance for the continuous time and the discrete time cases.

\end{document}